\documentclass[10pt]{iopart}
\usepackage{amsfonts}
\usepackage{bm}
\usepackage{amssymb}
\usepackage{graphicx}   
\usepackage{color}      
\usepackage{comment}
\usepackage{cite}

\newcommand{\im}{\textrm{Im}}

\newcommand{\SC}{\textrm{SC}}

\begin{document}

\title{Effects of non-uniform distributions of gain and loss in photonic crystals}

\author{Alexander Cerjan and Shanhui Fan}
\ead{acerjan@stanford.edu $\textrm{and}$ shanhui@stanford.edu}
\address{Department of Electrical Engineering, and Ginzton Laboratory, Stanford  University,  Stanford,  California  94305,  USA}

\date{\today}

\begin{abstract}
We present a $\mathbf{k} \cdot \mathbf{p}$ theory of photonic crystals containing gain and loss 
in which the gain and loss are added to separate primitive cells of the underlying Hermitian system,
thereby creating a supercell photonic crystal. We show that the supercell bands of this system
can merge outward from the degenerate contour formed from folding the bands of the underlying Hermitian system 
into the supercell Brillouin zone, but that other accidental degeneracies in the band structure of the underlying Hermitian system 
do not yield band merging behavior.
Finally, we show that the modal
coupling matrix in PhCs with balanced gain and loss is trace-less, and thus the imaginary components of
the eigenvalues can only move relative to one another as the strength of the gain and loss is varied, without any collective motion.
\end{abstract}


\section{Introduction}

Nature provides only a few fundamental ingredients for designing photonic devices, changing the phase of a
signal through shifts in the refractive index, and altering the amplitude of a signal using gain and loss. Exploiting index
contrast has yielded enormous advances in control of the flow of light, 
and is a central feature of modern device design \cite{joannopoulos}. In contrast, gain and loss have typically been viewed
primarily as either a requirement for light generation or amplification, or as something to be strictly
avoided, respectively. However, recently there has been substantial interest in the potential
benefits and unique behaviors which are possible in systems containing patterned gain and/or loss.
The initial interests on these systems are generated by the physics associated with parity-time ($\mathcal{PT}$) symmetry, in which
the eigenvalues of the system form complex conjugate pairs after coalescing at an exceptional point \cite{bender_pt-symmetric_1999,bender_complex_2002,musslimani_optical_2008,makris_beam_2008,klaiman_visualization_2008,longhi_bloch_2009,guo_observation_2009,makris_pt-symmetric_2010,ruter_observation_2010,chong_pt-symmetry_2011,ge_conservation_2012,hodaei_parity-time_symmetric_2014,lin_unidirectional_2011,regensburger_parity-time_2012,feng_experimental_2013,peng_parity-time-symmetric_2014,chang_parity-time_2014,liertzer_pump-induced_2012,brandstetter_reversing_2014,peng_loss-induced_2014,lin_enhanced_2016,pick_densityEP_arxiv}. 
Subsequently, these significant developments in $\mathcal{PT}$ symmetric devices have also led to the
discovery of other types of systems whose eigenvalues come in complex conjugate
pairs \cite{cannata_schrodinger_1998,miri_supersymmetry-generated_2013,tsoy_stable_2014,makris_constant_2015,nixon_all-real_2016}, 
or exhibit counterintuitive phenomena \cite{cerjan_eigenvalue_arxiv}.

As an important class of systems exhibiting patterned gain and loss, in the past few years the properties
of multidimensional photonic crystals (PhCs) with gain and loss have been considered, and 
PhCs with $\mathcal{PT}$ symmetry have been shown to possess an array of intriguing 
properties. 
$\mathcal{PT}$ symmetric periodic systems exhibit band merging
behaviors which leads to the formation of rings and contours of exceptional points \cite{makris_beam_2008,szameit_pt-symmetry_2011,ramezani_exceptional-point_2012,cerjan_zipping_2016},
a $\mathcal{PT}$ superprism effect, an all-angle supercollimation effect, and
the opening of band-gaps \cite{mock_pt_phc_2016}. Increasing the strength of the gain and loss in the system
can also result in the coalescence of exceptional points \cite{ding_coalescence_2015}. Band merging effects and exceptional rings
have also been observed in passive PhC slabs containing an accidental degeneracy \cite{zhen_spawning_2015}, in which
the non-Hermitian behaviors are acquired through the radiative losses of the resonances of the PhC slab. 
In some of these previously studied systems \cite{szameit_pt-symmetry_2011,zhen_spawning_2015,ge_lieb_lattice_2015,cerjan_zipping_2016}, degeneracies in the underlying Hermitian system have
been used to decrease the experimental requirements for observing the aforementioned behaviors.
This can be accomplished through the judicious addition of the gain and loss resulting in the coupling of the modes of the underlying systems forming such
a degeneracy, leading to the eigenvalues becoming complex for any non-zero strength of the gain and loss \cite{ge_parity-time_2014}.
Overall, these studies of PhCs with gain and loss have the potential to dramatically expand the capabilities of photonic circuits and devices.

In this article, we develop a theory of supercell PhCs in which the gain and loss is balanced within each
supercell, but are not necessarily $\mathcal{PT}$ symmetric. Using this analytic theory,
we show that band merging originating from a degenerate contour formed by supercell band folding, 
found in $\mathcal{PT}$ symmetric PhCs,
also persists for this more general class of systems, but does not occur for other accidental degeneracies in the band structure of the underlying Hermitian system.
In particular, the inability for a supercell PhC with balanced gain and loss to couple accidental degeneracies
allows for the formation of photonic band gaps via gain/loss modulation. Finally, we show that the modal
coupling matrix in PhCs with balanced gain and loss is trace-less, and thus the imaginary components of
the eigenvalues can only move relative to one another, without any collective motion.

The remainder of this article is organized as follows. In section \ref{sec:two} we review the
theory of supercell PhCs, and use a $\mathbf{k} \cdot \mathbf{p}$ theory to illuminate the
role of degeneracies in such systems. Section \ref{sec:three} provides a discussion of the properties
of $\mathcal{PT}$ symmetric PhCs. Then, in section \ref{sec:four}, we expand the theory of
supercell PhCs to incorporate systems which have balanced gain and loss, but are not necessarily
$\mathcal{PT}$ symmetric. Finally, some concluding remarks are given in section \ref{sec:conc}.

\section{Review of supercell photonic crystals \label{sec:two}}

\begin{figure}[t!]
\centering
\includegraphics[width=0.7\textwidth]{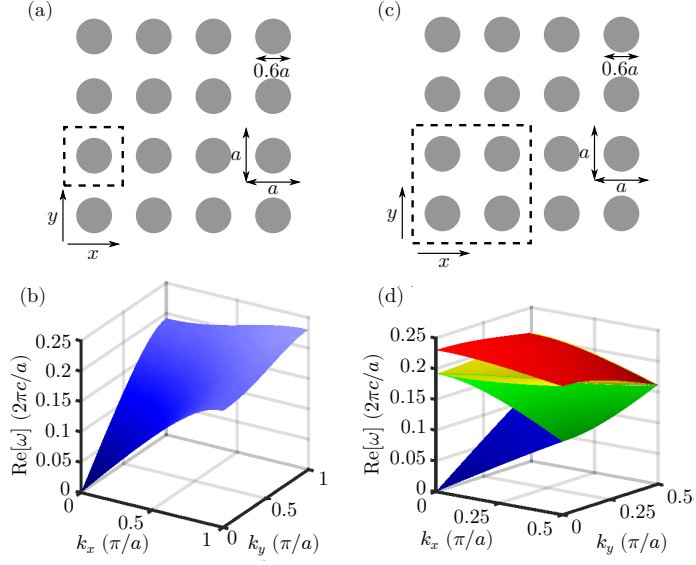}
\caption{(a) Schematic of a two-dimensional PhCs formed of dielectric circular rods, $\varepsilon = 12$, with radius
$l = 0.3a$, embedded in air, $\varepsilon_{\textrm{air}} = 1$, where $a$ is the lattice constant of the primitive cell.
The unit cell considered is denoted by the black dashed square.
(b) The first TM band of the PhC from (a) is plotted in the primitive Brillouin zone. (c),(d) Schematic and first primitive cell band of the same PhC as (a),
but evaluated instead over a supercell containing four dielectric rods. Because of this, the lone band in (b) is seen to be folded
into four constituent bands in (d) as the size of the supercell Brillouin zone is reduced.
\label{fig:1}}
\end{figure}

\begin{figure}[t!]
\centering
\includegraphics[width=0.99\textwidth]{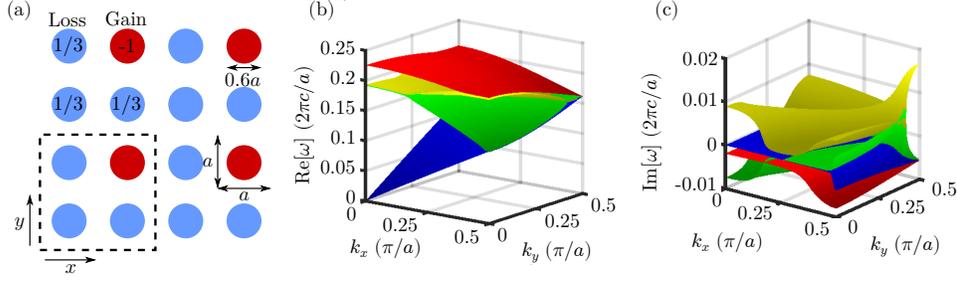}
\caption{(a) Schematic of a two-dimensional PhCs formed of dielectric circular rods, $\varepsilon = 12$, with radius
$l = 0.3a$, embedded in air, $\varepsilon_{\textrm{air}} = 1$, where $a$ is the lattice constant of the primitive cell.
The unit cell considered is denoted by the black dashed square. The red rods contain gain, with a relative strength of
$1$, while the blue rods contain loss, with a relative strength of $1/3$. As such, the gain and loss are balanced within
each unit cell. (b),(c) The real and imaginary parts of the
first four TM bands of this PhC when $\tau = 3$.
\label{fig:1b}}
\end{figure}

The fundamental goal of designing a `supercell' photonic crystal is to create
contours of degeneracies within the Brillouin zone whose states can then be
immediately coupled through the non-uniform application of gain and loss. 
To create such degenerate contours in the band structure, we consider photonic crystals
in which the application of gain and loss expands the primitive cell of the system,
which results in each band of the original primitive cell of the underlying Hermitian system
being folded into multiple bands in the reduced Brillouin zone of the non-Hermitian system. 
An example of this folding process is shown in Fig.~\ref{fig:1}. The first
transverse magnetic (TM) band of the Hermitian system, Fig.~\ref{fig:1}(b), is
folded into four bands when the primitive cell of the system is increased to
contain four elements, as shown in Fig.~\ref{fig:1}(d). (TM modes have their electric field
oriented along the $z$-axis.) In this case, the folding
process has produced degenerate contours between the first and second, as well
as the third and fourth bands along the $X$-$M$ and $Y$-$M$ directions, in addition to
a degenerate contour between the second and third bands along the $\Gamma$-$M$ direction.
As can be seen, this folded band structure of the Hermitian system contains all of the same information
as the unfolded band. However, when gain and loss are added such
that this fictitious primitive cell of the Hermitian system becomes the true primitive
cell of the non-Hermitian system, such as the distribution of gain and loss shown in Fig.~\ref{fig:1b}(a), the modes comprising the
degenerate contours can couple, as shown in Fig.~\ref{fig:1b}(b) and (c).
For semantic convenience, we will henceforth refer to the primitive
cell of the non-Hermitian system as the `supercell,' and reserve `primitive cell'
to refer to the primitive cell of the underlying Hermitian system.

To understand the coupling process which can occur at these degenerate contours,
one must calculate the band structure of the PhC, which is defined by the solutions of
\begin{equation}
\left[ \nabla \times \nabla \times - \left(\varepsilon(\mathbf{x}) + i \tau g(\mathbf{x}) \right) \frac{\omega_m^2(\mathbf{k})}{c^2} \right] \mathbf{E}_{m\mathbf{k}}(\mathbf{x}) = 0, \label{eq:phcWave}
\end{equation}
in which $\mathbf{E}_{m\mathbf{k}}(\mathbf{x})$ is the mode profile of the $m$th band of the supercell non-Hermitian system
with wavevector $\mathbf{k}$ and frequency $\omega_m(\mathbf{k})$. The dielectric of the underlying Hermitian system is given
by $\varepsilon(\mathbf{x})$, which is assumed to be periodic over the set of lattice vectors which define the primitive
cell, $\{ \mathbf{a}_i \}$, such that $\varepsilon(\mathbf{x} + \mathbf{a}_i) = \varepsilon(\mathbf{x})$. The distribution
of the additional gain and loss is given by $g(\mathbf{x})$, while $\tau \ge 0$ represents the strength of the added gain
and loss. Regions of gain have $g(\mathbf{x}) < 0$. The supercell is assumed to have lattice vectors $\{ \mathbf{A}_i \}$,
which are comprised of integer multiples of the primitive cell lattice vectors, $\mathbf{A}_i = \sum_j n_{ij} \mathbf{a}_j$, 
such that $g(\mathbf{x} + \mathbf{A}_i) = g(\mathbf{x})$. Given the periodicity of the structure, the mode
profiles obey the supercell translational symmetry,
\begin{equation}
\mathbf{E}_{m\mathbf{k}}(\mathbf{x} + \mathbf{A}_i) = e^{i \mathbf{k} \cdot \mathbf{A}_i} \mathbf{E}_{m\mathbf{k}}(\mathbf{x}).
\end{equation}

We now use a $\mathbf{k} \cdot \mathbf{p}$ theory \cite{johnson_k.p_1993,johnson_theory_1994,sipe_vector_2000} to elucidate the effect
of changing the strength of the gain and loss from the properties of the underlying Hermitian system.
To do so, we expand the mode profiles of the non-Hermitian system at wavevector $\mathbf{k}$ over the 
basis of states of the underlying Hermitian system at wavevector $\mathbf{k}_0$,
\begin{equation}
\mathbf{E}_{m\mathbf{k}}(\mathbf{x}) = \sum_l C_{ml}(\mathbf{k}) e^{i(\mathbf{k}-\mathbf{k}_0)\cdot \mathbf{x}} \mathbf{E}_{l\mathbf{k}_0}^{(0)}(\mathbf{x}). \label{eq:eExp}
\end{equation}
Here, $\mathbf{E}_{l\mathbf{k}}^{(0)}(\mathbf{x})$ has frequency $\omega_l^{(0)}(\mathbf{k})$ and satisfies equation (\ref{eq:phcWave}) with $\tau = 0$, and
$C_{ml}$ are the complex modal expansion coefficients. Using the mode profiles of the underlying Hermitian system
provides the additional benefit of allowing us to normalize the wave functions in the usual manner,
\begin{equation}
\int_{\SC} \varepsilon(\mathbf{x}) \left(\mathbf{E}_{l'\mathbf{k}}^{(0)}(\mathbf{x}) \right)^* \cdot \mathbf{E}_{l\mathbf{k}'}^{(0)}(\mathbf{x}) d\mathbf{x} = \delta_{ll'} \delta(\mathbf{k}-\mathbf{k}'), \label{eq:norm}
\end{equation}
where $\textrm{SC}$ denotes performing the integral over the supercell.

By multiplying equation (\ref{eq:phcWave}) through by
$(\mathbf{E}_{l'\mathbf{k}}^{(0)}(\mathbf{x}))^*$ and integrating over the supercell, we find
the equation
\begin{equation}
\sum_{l} \left[\frac{\omega_m^2(\mathbf{k})}{c^2} \delta_{l'l} - \Omega_{l'l}(\mathbf{k},\mathbf{k}_0) + i \tau \frac{\omega_m^2(\mathbf{k})}{c^2} G_{l'l}(\mathbf{k}) \right]C_{ml}(\mathbf{k}) = 0,
\end{equation}
which can be written as an explicit matrix equation as
\begin{equation}
\left[\frac{\omega_m^2(\mathbf{k})}{c^2} I - \Omega(\mathbf{k},\mathbf{k}_0) + i \tau \frac{\omega_m^2(\mathbf{k})}{c^2} G(\mathbf{k}) \right]C_{m}(\mathbf{k}) = 0, \label{eq:phcMat}
\end{equation}
in which $C_m$ is a column vector with elements $C_{ml}$. For the ease of the following analysis within Sec.~\ref{sec:two},
we have specialized to 2D TM bands and hence the electric field becomes a scalar, but a full vectorial treatment is straightforward \cite{sipe_vector_2000}.
In Eq.~(\ref{eq:phcMat}), $G$ contains the effects of modal coupling through the gain and loss
\begin{equation}
G_{l'l} = \int_{\SC} g(\mathbf{x}) \left(E_{l'\mathbf{k}}^{(0)}(\mathbf{x})\right)^* E_{l\mathbf{k}'}^{(0)}(\mathbf{x}) d\mathbf{x}, \label{eq:G1}
\end{equation}
and in general $G$ will have both diagonal and off-diagonal elements.
Likewise, $\Omega(\mathbf{k},\mathbf{k}_0)$ contains information about the underlying Hermitian system,
\begin{equation}
\Omega_{l'l}(\mathbf{k},\mathbf{k}_0) = \left(\frac{\omega_{l}^{(0)}(\mathbf{k}_0)}{c}\right)^2 \delta_{l'l} - \mathbf{s} \cdot \mathbf{P}_{l'l} + s^2 Q_{l'l},
\end{equation}
in which $\mathbf{s} = \mathbf{k}-\mathbf{k}_0$. Here, the frequencies of the underlying Hermitian system are contained
along the diagonal of $\Omega$, while the effects of choosing $\mathbf{k} \ne \mathbf{k}_0$ are given by
\begin{eqnarray}
\mathbf{P}_{l'l} &= 2i\int_{\textrm{SC}} \left(E_{l'\mathbf{k}_0}^{(0)}(\mathbf{x}) \right)^* \nabla E_{l\mathbf{k}_0}^{(0)}(\mathbf{x}) d\mathbf{x}, \\
Q_{l'l} &= \int_{\textrm{SC}} \left(E_{l'\mathbf{k}_0}^{(0)}(\mathbf{x}) \right)^* E_{l\mathbf{k}_0}^{(0)}(\mathbf{x})  d\mathbf{x}.
\end{eqnarray}

Equation (\ref{eq:phcMat}) is an exact restatement of Eq.~(\ref{eq:phcWave}), but benefits from
the explicit isolation of the effects of the strength of the gain and loss from the properties of
both the underlying Hermitian system, as well as the distribution of the non-Hermitian material.
Moreover, there is an advantage to using the modes of the underlying Hermitian system as
a basis set: these modes can be chosen to obey an additional `hidden' translational symmetry due
to the primitive cell translational symmetry of $\varepsilon(\mathbf{x})$ \cite{allen_recovering_2013}.
As the supercell with $\tau = 0$ is an exact $N$-fold copy of the primitive cell, there are exactly $N$
vectors $\{ \mathbf{c}_j \}$ which are integer multiples
of the primitive cell lattice vectors, such that translating the primitive cell along these $N$ vectors generates the supercell. 
Likewise, there are $N$ vectors, $\{ \mathbf{L}_j \}$,
which are integer multiples of the supercell reciprocal lattice vectors, 
such that translating the supercell Brillouin zone along these $N$ vectors generates the primitive Brillouin zone. 
The band folding that results from considering a supercell of the
underlying Hermitian PhC, such as the band structure shown in figure \ref{fig:1}(d), requires that each band in the primitive Brillouin zone
is folded into $N$ bands of the supercell. Thus, we can relabel the mode profiles as $l = (\nu,j)$, where
the $l$th band of the supercell is the $j$th fold of the $\nu$th band in the primitive Brillouin zone.
The hidden translational symmetry of the supercell Hermitian system then allows for the states 
$\mathbf{E}_{(\nu,j)\mathbf{k}}^{(0)}$ to be chosen such that \cite{allen_recovering_2013}
\begin{equation}
\mathbf{E}_{(\nu,j)\mathbf{k}}^{(0)}(\mathbf{x}+\mathbf{a}_i) = e^{i(\mathbf{k}+\mathbf{L}_j)\cdot \mathbf{a}_i} \mathbf{E}_{(\nu,j)\mathbf{k}}^{(0)}(\mathbf{x}). \label{eq:hidden}
\end{equation}
This hidden translational symmetry of the underlying Hermitian system
imposes strong selection rules upon the coupling matrices \cite{cerjan_zipping_2016},
\begin{eqnarray}
\mathbf{P}_{(\nu',j')(\nu,j)} &\sim \delta_{jj'} \label{eq:Psel}\\
Q_{(\nu',j')(\nu,j)} &\sim \delta_{jj'}. \label{eq:Qsel}
\end{eqnarray}

\section{Parity-time symmetric PhCs and exceptional contours \label{sec:three}}

Up through this point in reviewing the theory of supercell PhCs, no requirements have been made upon
the distribution of gain and loss beyond the expansion of the primitive cell of the system.
However, to finish the analysis of why degenerate contours are important features in supercell PhCs, we
now briefly specialize to $\mathcal{PT}$ symmetric systems.

In $\mathcal{PT}$ symmetric systems, the distribution of gain and loss is odd about a chosen
axis. This yields an additional selection rule on the modal coupling matrix $G$ \cite{cerjan_zipping_2016},
\begin{equation}
\mathcal{P}g(\mathbf{x}) = -g(\mathbf{x}) \;\;\; \Rightarrow \;\;\; G_{(\nu,j)(\nu,j)} = 0.
\end{equation}
Thus, in the neighborhood of a two-fold degenerate contour, $\mathbf{s} \cdot \mathbf{A}_i/2\pi \ll 1$, the system can be approximated as a 
two-level system, and Eq.~(\ref{eq:phcMat}) can be rewritten as
\begin{equation}
\frac{\omega_n^2(\mathbf{k})}{c^2} \left[ 
\begin{array}{cc}
1  & i\tau G_{12} \\
i\tau G_{21} & 1 
\end{array} \right] C_n = \Omega(\mathbf{k},\mathbf{k}_0) C_n, \label{eq:2lv}
\end{equation}
in which the labels $1$ and $2$ refer to the two supercell bands forming the degenerate contour.
As the two-level approximation of $\Omega$ is diagonal from equations (\ref{eq:Psel}) and (\ref{eq:Qsel}), the only coupling
between the two levels in Eq.~(\ref{eq:2lv}) comes from the application of the gain and loss.

\begin{figure}[t!]
\centering
\includegraphics[width=0.99\textwidth]{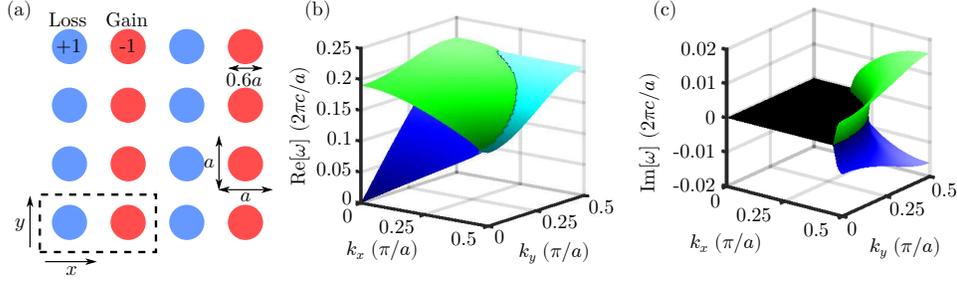}
\caption{(a) Schematic of a two-dimensional PhCs formed of dielectric circular rods, $\varepsilon = 12$, with radius
$l = 0.3a$, embedded in air, $\varepsilon_{\textrm{air}} = 1$, where $a$ is the lattice constant of the primitive cell.
The unit cell considered is denoted by the black dashed square. The red and blue rods contain gain and loss, respectively,
with equal strength. (b),(c) The real and imaginary parts of the first two TM bands of this PhC when $\tau = 2$.
The cyan region in (b) denotes where the bands have coalesced, and the constituent frequencies form complex conjugate pairs.
\label{fig:pt}}
\end{figure}

In general, for $\mathcal{PT}$ symmetric systems, as gain and loss are added to the system,
every pair of frequencies with the same $\mathbf{k}$ will begin to merge together until coalescing at an exceptional point,
beyond which the two frequencies will form complex conjugate pairs. 
In particular, if we choose the origin for the $\mathbf{k} \cdot \mathbf{p}$ expansion, $\mathbf{k}_0$, to
lie along the degenerate contour, $\omega_1^{(0)}(\mathbf{k}_0) = \omega_2^{(0)}(\mathbf{k}_0)$, then
we can see that any non-zero amount of gain and loss will lead frequencies along the degenerate
contour to form complex conjugate pairs. Furthermore, as the bands separate as $\mathbf{s}$ is
increased away from the degenerate contour, increasing amounts of gain and loss are needed to
make these frequencies coalesce. Thus, it can be shown that within the two-level approximation, 
the strength of the gain/loss required for the frequencies at a particular location in wavevector space 
to coalesce, $\tau_{\textrm{th}}$, is dependent upon the distance of that location to the nearest point on the degenerate
contour, $s_\perp$, to leading order,
\begin{equation}
\tau_{\textrm{th}} \approx \left| \frac{s_\perp P_{\perp}}
{|G_{12}|\left(\frac{\omega^{(0)}(\mathbf{k}_0)}{c}\right)^2} \right| + O(s^2). \label{eq:tauPred}
\end{equation}
Thus, as gain and loss are added to a periodic system in a $\mathcal{PT}$ symmetric distribution,
the degenerate contours undergo threshold-less $\mathcal{PT}$ transitions, while at
locations further away from the degenerate contour in wavevector space, the frequencies
coalesce for increasing $\tau$. This band merging process is shown in figure \ref{fig:pt}, which uses the
same underlying Hermitian system as is shown in figure \ref{fig:1}(a), but with equal gain and loss
added to alternating columns of dielectric rods so as to form a $\mathcal{PT}$ symmetric PhC. The boundary
between the band merged and unmerged regions forms a contour comprised entirely of exceptional points,
i.e.\ an exceptional contour, which originates immediately adjacent to the degenerate contour for small
but non-zero $\tau$, and then moves away from the degenerate contour as $\tau$ is increased, leaving
a region of frequencies which form complex conjugate pairs in its wake. 
Finally, to first order in $\mathbf{s}$,
the real part of the frequencies within the band merged region become equal to the frequency of the 
nearest point on the degenerate contour, $\omega^{(0)}(\mathbf{k}_0)$, creating flat isofrequency contours
in wavevector space. However, as we will show in the next section, this is in fact a more general feature
of PhCs with balanced gain and loss.

\section{Systems with balanced gain and loss \label{sec:four}}

Previously, the theory of supercell PhCs, developed in section \ref{sec:two}, has been applied to understand the
behavior of the band structure of $\mathcal{PT}$ symmetric PhCs, as discussed in section \ref{sec:three}. In this section,
we instead apply this theory to elucidate the band structure behavior of supercell PhC systems in which the gain and loss
is balanced within each supercell, but are not necessarily $\mathcal{PT}$ symmetric. 
Using this analytic theory,
we show that band merging originating from a degenerate contour formed by supercell band folding, 
discussed in section \ref{sec:three} for $\mathcal{PT}$ symmetric PhCs,
also persists for this more general class of systems, but does not occur for other accidental degeneracies in the band structure of the underlying Hermitian system.
In particular, the inability for a supercell PhC with balanced gain and loss to couple accidental degeneracies
allows for the formation of photonic band gaps via gain/loss modulation. Finally, we show that the modal
coupling matrix $G$ in PhCs with balanced gain and loss is trace-less, and thus the imaginary components of
the eigenvalues can only move relative to one another, without any collective motion.

As an analytic approach, we assume that the gain and loss are simply added to separate primitive cells
of the underlying system, and need not be distributed in a $\mathcal{PT}$ symmetric manner, such that
\begin{equation}
g(\mathbf{x}) = g_n \varepsilon(\mathbf{x}), \qquad \mathbf{x} \in \textrm{PC}_n, \label{eq:gn}
\end{equation}
in which $g_n$ represents the gain or loss added to the $n$th primitive cell, $\textrm{PC}_n$.
Although it is possible to conceive of PhCs with balanced gain and loss which do not satisfy this criteria,
adding gain and loss to separate elements of the PhC represents the most experimentally feasible construction
of such a system.
Using equation (\ref{eq:gn}) we rearrange equation (\ref{eq:G1}) as
\begin{equation}
G_{(\nu',j')(\nu,j)} = \sum_{n=1}^N g_n \int_{\textrm{PC}_n} \varepsilon(\mathbf{x}) \left(\mathbf{E}_{(\nu',j')\mathbf{k}}^{(0)}(\mathbf{x}) \right)^* \cdot \mathbf{E}_{(\nu,j)\mathbf{k}}^{(0)}(\mathbf{x}) d\mathbf{x}, \label{eq:GG}
\end{equation}
where now the integrals are performed over each of the $N$ primitive cells (PCs) which comprise the supercell.

Equation (\ref{eq:GG}) can be simplified by using the hidden translational symmetry of the states $\mathbf{E}_{(\nu,j)\mathbf{k}}^{(0)}$.
Each integral over a constituent primitive cell is related to the fundamental
primitive cell by $\mathbf{x} = \mathbf{x}' + \mathbf{c}_n$, in which $\mathbf{x} \in \textrm{PC}_n$, 
$\mathbf{x}' \in \textrm{PC}_1$, and where the fundamental primitive cell, $\textrm{PC}_1$, is defined as the primitive cell with
$\mathbf{c}_1 = 0$. Thus, equation (\ref{eq:GG}) can be rewritten as,
\begin{equation}
G_{(\nu',j')(\nu,j)} = \left( \int_{\textrm{PC}_1} \varepsilon(\mathbf{x}) \left(\mathbf{E}_{(\nu',j')\mathbf{k}}^{(0)}(\mathbf{x}) \right)^* \cdot \mathbf{E}_{(\nu,j)\mathbf{k}}^{(0)}(\mathbf{x}) d\mathbf{x} \right)
\sum_{n=1}^N g_n e^{i(\mathbf{L}_{j}-\mathbf{L}_{j'})\cdot \mathbf{c}_n}.
\end{equation}
Here, the remaining integral is none other than the primitive cell orthogonality relationship,
\begin{equation}
\int_{\textrm{PC}_1} \varepsilon(\mathbf{x}) \left(\mathbf{E}_{(\nu',j')\mathbf{k}}^{(0)}(\mathbf{x}) \right)^* \cdot \mathbf{E}_{(\nu,j)\mathbf{k}}^{(0)}(\mathbf{x}) d\mathbf{x} = \frac{\delta_{\nu \nu'}}{N}, \label{eq:PCnorm}
\end{equation}
in which the factor of $1/N$ is required for chosen normalization of the supercell states in equation (\ref{eq:norm}).
Thus, for any supercell PhC, the modal coupling due to non-Hermitian material can be written as
\begin{equation}
G_{(\nu',j')(\nu,j)} = \frac{\delta_{\nu \nu'}}{N} \sum_{n=1}^N g_n e^{i(\mathbf{L}_{j}-\mathbf{L}_{j'})\cdot \mathbf{c}_n}. \label{eq:GF}
\end{equation}

\begin{figure}[t!]
\centering
\includegraphics[width=0.8\textwidth]{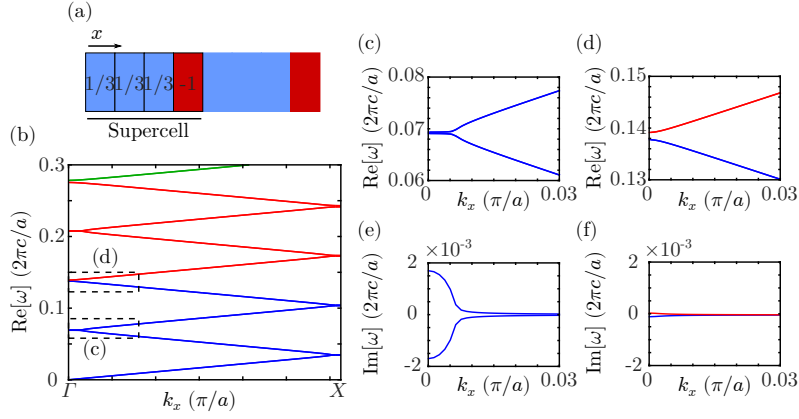}
\caption{(a) Schematic of a one-dimensional PhC in which the underlying Hermitian system is a uniform
dielectric, $\varepsilon = 13$, while the gain and loss are patterned on top of the system such that
the lossy region is three times longer than the gain region, but also three times weaker, so that the
gain and loss are balanced within each supercell. (b) Real parts of the bands
for this PhC when $\tau = 3$. (c),(d) Show a zoomed-in portion of the real part of the frequencies, as indicated
by the dashed boxes in (b). (e),(f) Imaginary parts of the bands for the same bands shown in (c) and (d),
respectively.
\label{fig:bragg}}
\end{figure}

This result has three important consequences. First, equation (\ref{eq:GF})
demonstrates that supercell PhCs satisfying the gain and loss patterning requirements of equation (\ref{eq:gn})
can never achieve coupling between different bands of the underlying Hermitian system.
Thus, within the two-level approximation of the supercell $\mathbf{k} \cdot \mathbf{p}$
theory, degeneracies which exist within the primitive Brillouin zone persist, and remain uncoupled, 
in the supercell band structure even as gain and loss are added. However, higher order coupling effects
are likely to lift such degeneracies, and thus have the potential to open a photonic band gap.
In particular, each band which is a part of such an uncoupled degeneracy is likely coupled to a different
band elsewhere in the supercell Brillouin zone, which can cause the entire band to shift slightly as
part of the band merging process as $\tau$ is increased.
Note that this first consequence does not require the gain and loss to be balanced.

An example of this process is shown in figure \ref{fig:bragg}, which
shows a photonic crystal formed of gain and loss patterned on top of a one-dimensional, uniform dielectric,
and in which the gain and loss are chosen to be balanced within each supercell. The folded band structure
of this system is shown in figure \ref{fig:bragg}(b), in which the different bands of the primitive Brillouin
zone are shown in different colors. As can be seen in figure \ref{fig:bragg}(c), two supercell bands which
correspond to the same primitive band can couple, which results in the flattening of the band near $\Gamma$, 
while the two supercell bands originating from different
primitive bands cannot, as observed in figure \ref{fig:bragg}(d), 
consistent with the general discussion given in the previous paragraph.
This effect has been observed by Mock in $\mathcal{PT}$ symmetric PhCs formed by patterning
gain and loss on top of uniform underlying Hermitian systems \cite{mock_pt_phc_2016},
but here we have proven that this effect is generally present in systems with patterned gain and loss,
and should also be observable in systems containing accidental
degeneracies \cite{huang_dirac_2011,chan_dirac_2012}.

\begin{figure}[t!]
\centering
\includegraphics[width=0.85\textwidth]{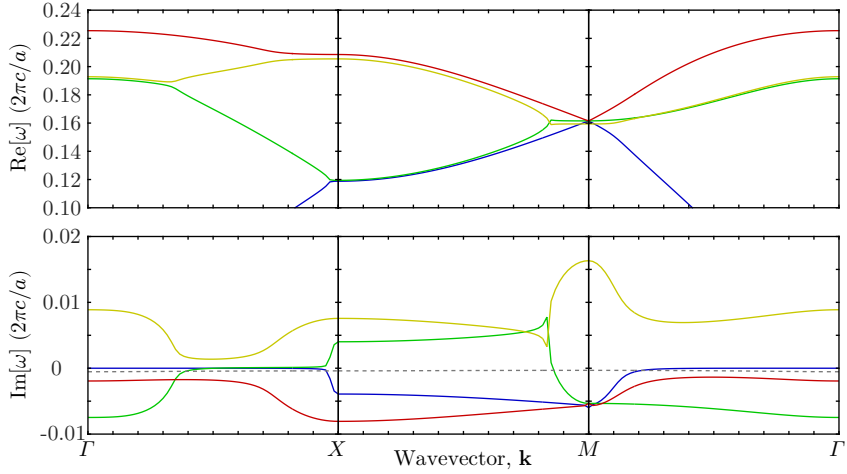}
\caption{Band structure for the PhC shown in Fig.~\ref{fig:1b}(a) around the irreducible zone boundary.
The real parts of the bands are shown in the top panel, while the imaginary parts of the bands are shown
in the bottom panel. The gray dashed line in the bottom panel plots the sum of the imaginary parts of the
four frequencies shown.
\label{fig:cut}}
\end{figure}

The second important consequence of equation (\ref{eq:GF}) is that if the gain and loss in the
system are balanced, then there is no self-coupling through the non-Hermitian material,
\begin{equation}
\sum_{n=1}^N g_n = 0 \;\;\; \Rightarrow \;\;\; G_{(\nu,j)(\nu,j)} = 0. \label{eq:bal}
\end{equation}
This means that in any supercell system with balanced gain and loss the matrix $G$ is trace-less, 
and moreover in the neighborhood of a two-fold degenerate contour the system can be approximately described by equation (\ref{eq:2lv}).
Thus, as the strength of the gain and loss is increased, the coupled bands will merge outward from the
degenerate contours, forming flat regions and complex conjugate frequency pairs, similar to
what is seen in a $\mathcal{PT}$ symmetric system, and given by equation (\ref{eq:tauPred}).
An example of this band merging process for a PhC with balanced gain and loss can be seen in figure \ref{fig:cut},
which shows the band structure along the irriducible Brillouin zone boundary for the same PhC shown in figure \ref{fig:1b}(a).
Here, near $X$ along the $\Gamma$-$X$ direction, two sets of coupled bands are seen to be in the process
of merging, forming nearly flat regions perpendicular to the associated degenerate contour, which runs along the $X$-$M$ direction, and possessing large imaginary 
components which are nearly complex conjugate pairs. Near $X$, the first (blue) band is merging with the second (green) band,
and likewise the third (yellow) band is merging with the fourth (red) band. 
Similarly, along the majority of the $\Gamma$-$M$ direction, the second (green) and third (yellow) bands
are seen to be merged, forming nearly complex conjugate pairs with large imaginary components along this direction.

In the vicinity of a two-fold degenerate line, the band merging behavior found in PhCs with balanced gain and loss
is similar to the band merging behavior observed in $\mathcal{PT}$ symmetric PhC systems as described in section \ref{sec:three}. 
The only difference between the coalescence observed near two-fold degenerate contours
in PhCs which have balanced gain and loss, but are not $\mathcal{PT}$ symmetric, and $\mathcal{PT}$ symmetric PhCs, is that
the coalescence in the former does not produce an exceptional contour, as the real parts of the frequencies are not
exactly equal. This can be understood as a consequence of the slight impedence mismatch that exists between the
gain and loss elements, as in non-$\mathcal{PT}$ symmetric systems $|\varepsilon_{gain}| \ne |\varepsilon_{loss}|$,
which breaks the exact symmetry that yields an exceptional point in $\mathcal{PT}$ symmetric systems.
Thus, although the two-level model in equation (\ref{eq:2lv}) predicts that PhCs with balanced gain and
loss should also exhibit exceptional contours, the two-level approximation in this case omits some weak coupling elements
of $G$. In contrast, no such omission of the elements of $G$ is required to construct a two-level model of 
the $\mathcal{PT}$ symmetric PhC shown in figure \ref{fig:pt}.

At this point, we should note that the distribution of gain and loss in the PhC shown in figure \ref{fig:1b}
is not exactly described by the condition specified in equation (\ref{eq:gn}), as gain and loss have not
been added to the air surrounding the dielectric rods. Nevertheless, equation (\ref{eq:gn}) provides an
excellent approximation for systems comprised of high- and low-dielectric regions in which gain and loss are
only added to the high-dielectric portions for two reasons.
First, the amount of gain/loss omitted from the low-dielectric region is already less than the amount in the high-dielectric 
region by the ratio $\varepsilon_{\textrm{low}}/\varepsilon_{\textrm{high}} \sim 0.1$.
And second, we expect the mode
profiles to be concentrated in the high-dielectric regions where they will be relatively unaffected by
the presence or absence of gain or loss in the low-dielectric regions.
As such, we expect the theory developed here to be applicable to most experimental systems.

The third consequence of equation (\ref{eq:GF}) is that the eigenvalues of systems with
balanced gain and loss can only move relative to one another, as their average is fixed.
One can observe this behavior in the bottom panel of figure \ref{fig:cut}, where the imaginary 
part of the eigenfrequencies for the four bands sum to near-zero for every $\mathbf{k}$.
To explain this behavior, 
we first note that the coupling matrix $G$ does not couple bands
which originate from different bands in the primitive Brillouin zone. As such, when $\mathbf{k} = \mathbf{k}_0$, equation (\ref{eq:phcMat})
is block diagonal, comprised of $N$-by-$N$ blocks formed from the supercell modes
corresponding to the same unfolded band in the primitive Brillouin zone. Thus,
without loss of generality, we restrict our analysis to a single $N$-by-$N$ block
of equation (\ref{eq:phcMat}). For small values
of $\tau$, we expect small deviations in the resulting band structure of the system,
and as such we expand the frequencies of the non-Hermitian system as $\omega_m^2(\mathbf{k}) = \bar{\omega}^2(\mathbf{k}) + \delta \lambda_m(\mathbf{k})$,
where $\bar{\omega}^2(\mathbf{k}) = c^2 \tr[\Omega(\mathbf{k},\mathbf{k})]/N$ is the average frequency
of the set of supercell bands under consideration. Thus, to first order in the small
quantities $\tau$ and $\delta \lambda_m$, equation (\ref{eq:phcMat}) can be approximated
as
\begin{equation}
\left[c^2 \Omega(\mathbf{k},\mathbf{k}) - \bar{\omega}^2(\mathbf{k}) I - i \tau \bar{\omega}^2(\mathbf{k}) G(\mathbf{k}) \right]C(\mathbf{k}) = C(\mathbf{k}) \delta \Lambda(\mathbf{k}), \label{eq:Mat2}
\end{equation}
in which $\delta \Lambda$ is the diagonal matrix of the eigenvalues $\delta \lambda_m$, and
$C$ is the matrix whose columns are the coefficient eigenvectors $C_m$.

The main advantage of equation (\ref{eq:Mat2}) is that it is an ordinary eigenvalue problem, 
rather than a generalized eigenvalue problem. As such, the sum of the eigenvalues
is given by the trace of the defining matrix,
\begin{equation}
\sum_m^N \delta \lambda_m(\mathbf{k}) = \tr\left[c^2 \Omega(\mathbf{k},\mathbf{k}) - \bar{\omega}^2(\mathbf{k}) I - i \tau \bar{\omega}^2(\mathbf{k}) G(\mathbf{k}) \right] = 0, \label{eq:lamSum}
\end{equation}
in which we have used the fact that $\tr[G]=0$ as shown in equation (\ref{eq:bal}).
Thus, the imaginary components of the eigenvalues
of the non-Hermitian system sum to zero within each set of folded supercell bands, $\sum_m^N \im[\delta \lambda_m^2(\mathbf{k})] = 0$.
While the proof above uses perturbation theory that is strictly speaking only for small $\tau$, equation (\ref{eq:lamSum})
is confirmed numerically in 
the bottom panel of figures \ref{fig:cut} even for $\tau = 3$.

The theoretical analysis presented here demonstrates a fundamental property of systems with balanced
gain and loss: if one mode acquires gain or loss, that gain or loss must be compensated by distributing an equal quantity 
the opposite non-Hermitian behavior among the remaining modes of the system.
This property is clearly observed in $\mathcal{PT}$ systems, in which the eigenvalues 
appear in complex conjugate pairs. However, in more general systems that simply
contain balanced gain and loss, more complex behaviors are possible. For example, consider
the $M$ point from the system studied in figures \ref{fig:1b} and \ref{fig:cut}. When
$\tau = 0$, $M$ is four-fold degenerate, and as $\tau$ is increased, all four of these modes
couple together. However, unlike in a $\mathcal{PT}$ PhC, these eigenvalues do not form
two complex conjugate pairs. Instead, a single mode becomes strongly amplifying, and the remaining
three modes become weakly absorbing. This behavior is potentially useful if one is able to couple
to a specific mode through cavity engineering, as it allows for one to enhance a desired signal while
simultaneously suppressing noise in the remaining channels of the system.

\section{Conclusion \label{sec:conc}}

In summary, we have presented a framework for understanding the eigenvalue dynamics of
PhCs containing balanced gain and loss as the strength of the gain and loss is increased.
We have shown that in such systems the imaginary portions of the frequencies of the bands move relative to
one another, but that there is no collective drift of the entire set of eigenvalues. In addition,
we have also demonstrated that degenerate contours are a critical element in the formation of
regions of the Brillouin zone with large imaginary frequency components. Although here we have focused
on developing a theory of PhCs with balanced gain and loss, these results extend to other similar systems,
both periodic and isolated.

\ack

This work was supported by an AFOSR MURI program (Grant
No.\ FA9550-12-1-0471), and an AFOSR project (Grant No.\ FA9550-16-1-0010). 



\section*{References}


\end{document}